# Analysis and Verification of Relation between Digitizer's Sampling Properties and Energy Resolution of HPGe Detectors
Jinfu Zhu, Tianhao Wang, Tao Xue, Liangjun Wei, Jingjun Wen, Lin Jiang, Jianmin Li

*Abstract*–The CDEX (China Dark matter Experiment) aims at detection of WIMPs (Weakly Interacting Massive Particles) and 0vbb (Neutrinoless double beta decay) of $^{76}$Ge. It now uses ~10 kg HPGe (High Purity Germanium) detectors in CJPL (China Jinping Underground Laboratory). The energy resolution of detectors is calculated via height spectrum of waveforms with 6-μs shaping time. It is necessary to know how sampling properties of a digitizer effect the energy resolution. This paper will present preliminary energy resolution results of waveforms at different sampling properties. The preliminary results show that the ENOB (effective number of bits) with 8.25-bit or better can meet the energy resolution @122keV of CDEX HPGe detectors. Based on the ADC (Analog-to-Digital Converter) quantized error theory, this paper will also make a quantitative analysis on energy resolution in CDEX HPGe detectors. It will provide guidance for ADC design in full-chain cryogenic readout electronics for HPGe detectors.

## I. Introduction

Dark matter and 0vbb (Neutrinoless double beta decay) have reached a very popular stage these years [1] [2] [3] [4]. One of operated experiments for dark matter is direct detection by HPGe (High Purity Germanium) detectors with high sensitivity and low radiation background. It focuses on the elastic scattering between WIMPs (Weakly Interacting Massive Particles) and HPGe detectors [5] [6], and annual modulation analysis of WIMPs [7]. 0vbb (Neutrinoless double beta decay) of $^{76}$Ge using ton-scale germanium detector array is also proposed [8]. The CDEX (China Dark matter Experiment) now deploys ~10 kg pPCGe (p-type Point Contact Germanium) detectors in CJPL (China Jinping Underground Laboratory) [9].

Energy resolution is an important index of germanium detectors. The energy resolution of CDEX HPGe detectors is calculated via the FWHM (Full Width at Half Maximum) of height spectrum. The height spectrum uses waveforms with 6-μs shaping time from JFET (Junction gate Field-Effect Transistor) preamplifiers. Those waveforms are digitized by 100 MSPS 14-bit ADCs (Analog-to-Digital Converters) [10].

Manuscript received October 22nd, 2020. This work was supported by the National Natural Science Foundation of China (U1865205).
The authors are with the Department of Engineering Physics, Tsinghua University, and also with the Key Laboratory of Particle & Radiation Imaging (Tsinghua University), Ministry of Education, Beijing, 100084, China (Corresponding author is Tao Xue, e-mail: xuetaothu@mail.tsinghua.edu.cn).

The full-chain cryogenic readout electronics for HPGe detectors is also designed by preamplifier and ADC ASICs (Application Specific Integrated Circuits) [11]. The cost and power consumption of ASICs is relevant with sampling properties. One question should be answered that what sampling properties of ADC can meet the requirement of energy resolution for HPGe detectors.

This paper is organized as follows. Section II introduces the method of quantitative analysis on energy resolution of CDEX HPGe detectors. Section III will introduce energy resolution results by different sampling properties. The conclusion and future work will be discussed in Section IV.

## II. Method

As shown in Fig. 1, for a waveform with 6-μs shaping time from JFET preamplifiers, its *height* is proportional to the deposited energy *E* of an incident particle. The waveform is digitized by ADC (Sampling rate is $F_S$, Full-scale is *FUS*, and total time window is *T*).

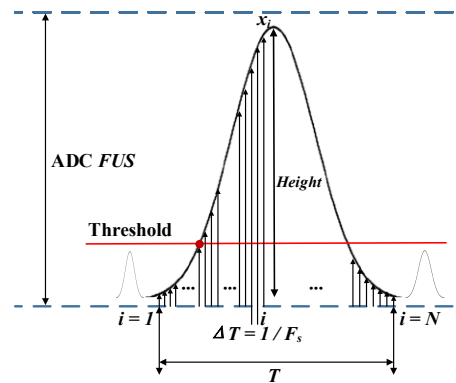

Fig. 1 Diagram of waveform digitization.

The Energy resolution $\eta$ can be defined as equation (1) and the typical gamma-ray energy spectrum ($^{57}$Co) is shown in Fig. 2. The analysis are based on a hypothesis that quantized noise from digitizer is uncorrelated with that from others, e.g. preamplifiers and detectors etc.

$$\eta = \Delta E = 2.355\sigma_E = 2.355\left(\sqrt{\sigma_{digitizer}^2 + \sigma_{other}^2}\right) \quad (1)$$

According to ADC quantized error format, the quantized noise from digitizer can be estimated as equation (2). *K* is the calibration parameter of energy spectrum and $\sigma_{height}$ is the MSE (mean square error) of pulse height. ENOB is the effective number of bits of ADC and *A* is the gain of digitizer.

Combining the equation (1) and (2), the relationship between energy resolution and digitizer performance can be written as equation (3).

$$\sigma^2_{digitizer} = \frac{(\sigma_{height} K)^2}{A^2} = \frac{K^2}{A^2}\left(\frac{FUS}{2^{ENOB}}\right)^2 \frac{1}{12} \quad (2)$$

$$\eta = 2.355\left(\sqrt{\frac{K^2}{A^2}\frac{FUS^2}{12}\left(\frac{1}{4^{ENOB}}\right)+\sigma^2_{other}}\right) \quad (3)$$

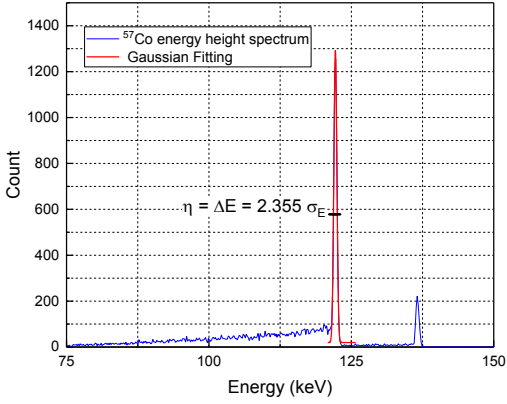

Fig. 2. The typical energy spectrum.

### III. EXPERIMENT AND RESULTS

The typical waveform with 6-μs shaping time generated by a 122 keV gamma-ray and its power spectrum by (Fast Fourier Transform) are show in Fig. 3. The waveform is digitized by 100 MSPS 14-bit ADC. The ENOB of ADC is ~11.25 bit using a standard sinusoidal signal according to IEEE 1241-2000 standard. It shows that the rise-time of waveform is several μs and the bandwidth of the waveform is less than 1 MHz. The sampling rate of ADC with 100 MSPS is efficient for digitization.

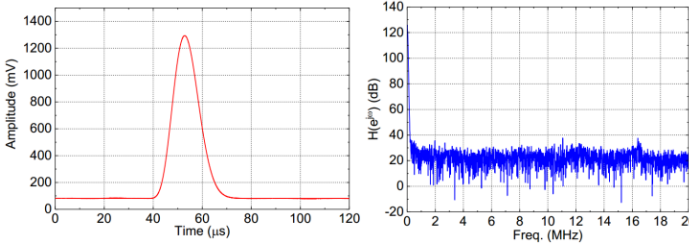

Fig. 3. The typical 122 keV waveform and its power spectrum.

The ENOB of ADC is declined by remove low data bits and energy resolution results at different ENOBs (ranged from 11.25 to 5.25 bit) are shown in Fig. 4. They are fitted well by the equation (3). The intrinsic energy resolution is estimated to be ~560 eV@122 keV if the ideal performance of ADC is achieved. It presents that the ENOB with 8.25-bit or better can meet the energy resolution @122keV of CDEX HPGe detectors.

### IV. SUMMARY AND FUTURE WORK

This paper analyzes the sampling properties of digitizer on energy resolution of HPGe Detectors. The method is presented based on quantized error of ADC. Experiments results based on CDEX HPGe detectors have verified the method. In the future, to improve the energy resolution more, other factors such as preamplifier and main amplifier will be researched.

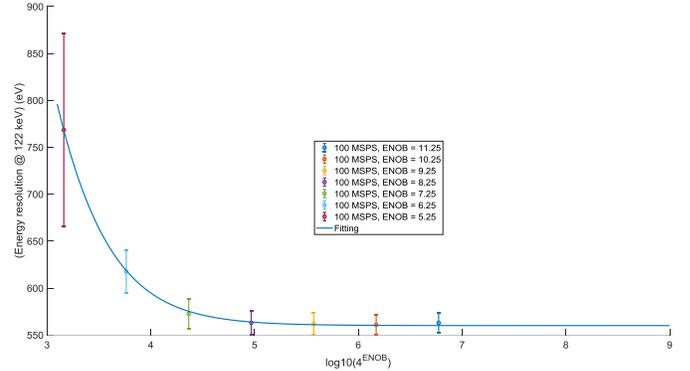

Fig. 4. Energy resolution results at different ENOBs and fitting result.

### ACKNOWLEDGEMENT

We would like to thank those who collaborated on the CDEX, and also thank Professor Hao Ma, Zhi Deng, and Guang Meng for their support and various discussions over the years at the Tsinghua University DEP (Department of Engineering Physics).

We are grateful for the patient help of Yu Xue, Wenping Xue, and Jianfeng Zhang. They are seasoned full-stack hardware technologist with wealth experience of solder and rework in the electronics workshop at DEP.